# Transition from physical to online shopping alternatives due to the COVID-19 pandemic - a case study of Italy and Sweden

Claudia Andruetto (corresponding author)

E-mail: andru@kth.se

Address: Drottning Kristinas Väg 40, 114 28, Stockholm, Sweden

Integrated Transport Research Lab, KTH Royal Institute of Technology, Sweden

Elisa Bin

Integrated Transport Research Lab, KTH Royal Institute of Technology, Sweden

Yusak Susilo

University of Natural Resources and Life Sciences (BOKU), Vienna

Anna Pernestål

Integrated Transport Research Lab, KTH Royal Institute of Technology, Sweden

# Abstract

Using 530 responses from an online questionnaire, this study aims to investigate the transition from physical to online shopping alternatives during the first wave of the COVID-19 pandemic at the individual level. The focus areas of the study are Sweden and Italy, two European countries that implemented contrasting prevention measures. This study analyses the impacts of the pandemic on the transition to online shopping activities, and identifies who among the respondents changed their shopping behaviour the most and how; and what the different shopping strategies are and who adopted them. Multivariate statistical analyses, including linear and binary logistic regressions and multinomial logit models, were used to analyse the dataset. In the analysis, the dataset was split between Italy and Sweden to take into account the contrasting prevention measures and the different social and economic backgrounds of the two countries; the results of this study confirm and highlight these differences. Moreover, the socio-demographic and household structures of the respondents were found to influence the amount and the direction of change in shopping behaviour during the first wave of the pandemic. The study also indicates some policies that can be implemented and/or further strengthened to increase the resilience of citizens in facing pandemics and to derive benefit from the behavioural changes that took place during the first wave of the pandemic.

# 1. Introduction

In 2020, the world faced the COVID-19 pandemic, after the virus first appeared in Wuhan, China at the end of 2019 (Arimura et al., 2020). Many governments worldwide responded to the situation with different lockdown strategies (e.g., travel bans, stay-at-home orders, restrictions on gatherings) designed to control the outbreak (Arimura et al., 2020; De Vos, 2020; Malik et al., 2020; Rahman et al., 2020). Significant reductions in the number of daily trips and public transport usage have been observed globally (Almlöf et al., 2020; Beck & Hensher, 2020; Bin et al., 2020; Lozzi et al., 2020; Molloy et al., 2020), but the pandemic had less effect on shopping trips (such as for grocery), which are considered essential (Parady et al., 2020).

Most of the analyses performed to date focus on the impacts of the pandemic and the lockdowns on passenger movements, public transport ridership, and traffic reduction (e.g., Arimura et al., 2020; Chan et al., 2020; De Vos, 2020; Kraemer et al., 2020; Kunzmann, 2020; Parady et al., 2020; Pepe et al., 2020; Warren & Skillman, 2020). Less attention has been given to the behavioural transition process that individuals have had to undergo in their daily lives: for example, the behavioural shift from physical to online activities due to the pandemic had a significant influence on shopping activities.

In this paper, the transition from physical to online shopping is studied through an online survey distributed during the first wave of the pandemic. In this study, the term "first wave" refers to the first spike in cases of COVID-19 observed in spring 2020, followed by a period of lockdown (Lockerd Maragakis, 2020). In particular, the study focuses on Italian and Swedish respondents. Italy and Sweden can be seen as two counterparts in Europe in terms of prevention measures, where Italy imposed a hard lockdown while Sweden's strategy was primarily based on recommendations from the government (Ministero della Salute, 2020; Regeringskansliet, 2020).

The main research question is: ***What changes in shopping behaviour took place during the first wave of the COVID-19 pandemic?*** In particular, the focus is on changes across different types of shopping and the differences between Italy and Sweden, as the governments of these two countries acted very differently in terms of policies and restrictions. In this paper, the behavioural changes among respondents from the two countries are examined and compared, taking into account the different restrictions in place and the socio-demographic of the respondents.

The research sub-questions are the following.

- **RQ1. Who changed their shopping behaviour the most, and how?**

  During the first wave of the pandemic, many people deviated from their usual behaviour, shifting towards online shopping and reducing the number of physical trips. This paper examines the characteristics of those who changed their behaviour the most.

- **RQ2. What shopping strategies were adopted during the first wave of the pandemic, and by whom?**

  During the lockdown period, people had varying capabilities to evaluate risks, adjust their daily routines to the imposed measures, and adopt alternative digital solutions for their daily shopping needs. Whilst such adjustments may come naturally to some groups in society, it may be more challenging for others. Therefore, there may have been some socio-demographic groups who did not feel safe enough to travel for shopping, but at the same time were not capable enough to employ digital alternatives during the first lockdown period. These groups risk being excluded from society and becoming incapable of fulfilling their daily needs.

The remainder of the paper is structured as follows. Section 2 comprises a literature review on COVID-19 and online shopping behaviours. In Section 3, the methodology is described. In Section 4, the data from the online questionnaire are presented, followed by descriptive and multivariate analyses. The paper concludes with the discussion and conclusion sections.

# 2. Literature review

COVID-19 has caused immense disruption to people's well-being and livelihoods. Over the last few months, there have been enormous efforts within the scientific community not only to develop a medical cure for the disease, but also to investigate how the disease has changed people's daily lives around the world. In terms of transport, various studies have been carried out that seek to characterise the pandemic period in terms of i) changes in urban mobility and behaviour (Arimura et al., 2020; Chan et al., 2020; Dahlberg et al., 2020; De Vos, 2020; Gao et al., 2020; Kavanagh et al., 2020; Kraemer et al., 2020; Kunzmann, 2020; Layer et al., 2020; Nielsen et al., 2020; Parady et al., 2020; Pepe et al., 2020; Quilty et al., 2020; Warren & Skillman, 2020, 2020); ii) impacts of lockdown strategies and social distancing measures (e.g. Malik et al., 2020; Rahman et al., 2020); and iii) changes in shopping activity (e.g. Hashem, 2020). Moreover, there is also literature focusing on how to characterise these changes and which section of the population has been impacted the most (e.g., Jay et al., 2020; Laurencin & McClinton, 2020; Yechezkel et al., 2020).

According to Rahman et al. (2020), adopting strict lockdown measures (seen as an effective way to maintain social distancing) significantly reduces overall mobility. The reduction is also associated with socioeconomic and institutional factors, such as average age, level of globalisation, and employment in service sectors (Rahman et al., 2020). According to data from Google and Apple, high-income countries experienced the greatest reduction in mobility (Apple, 2020; Google, 2020; Lozzi et al., 2020). Globally, public transport usage reached a low point of -76% in April 2020, while driving fell to -65% and walking to -67% (Apple, 2020; Lozzi et al., 2020). In Sweden, there was a 64% increase of the population in residential areas during working hours, a 38% decrease of maximum trip length, and a 33% average decrease of daytime presence in industrial and commercial areas (Dahlberg et al., 2020). In Italy, there was a reduction of 50% in the total number of trips between provinces (Pepe et al., 2020). In the following sections, the authors provide a brief overview of the main findings in terms of COVID-19 policies and transition behaviour to online shopping activities.

## 2.1 COVID-19 policies

Following the global outbreak of COVID-19 generated by the novel human coronavirus SARS-CoV-2 at the end of 2019 and the beginning of 2020, countries across the world took measures in order to reduce the spread of the virus, or at least to slow it down, in order to better cope with public health and better manage its limited resources (Musselwhite et al., 2020; WHO, 2020). Human-to-human transmission of SARS-CoV-2 has been described as having an incubation period between two and ten days, facilitating its spread via droplets, contaminated hands, or surfaces (Kampf et al., 2020). While knowledge about the SARS-CoV-2 virus was limited at the start of the pandemic, many public health researchers had experience with other infectious diseases that share some characteristics with the virus.

To help countries navigate the challenges presented by the pandemic, the World Health Organization (WHO, 2020) provided operational planning guidelines for balancing the demands of responding directly to COVID-19 while maintaining essential health-service delivery and mitigating the risk of system collapse. This includes prioritising the continuity of essential services delivery and making strategic shifts to ensure that increasingly limited resources provide maximum benefit for the population, whenever is needed. At the same time, travellers were also asked to comply with the highest standard in precautions, especially in hygiene practices, and wear personal protective equipment (WHO, 2020).

Whilst the prevention measures seem straightforward, for various other reasons, the implementation of those measures has varied across the world. For example, countries have differed in their implementation of social distancing measures, which aim to limit social contact and therefore prevent the spread of the virus. Some (including Italy) have enforced hard lockdowns, while others (including Sweden) have been less stringent (De Vos, 2020). The measures have been applied for different periods of time, and it is expected that new waves will result in changes to the measures (De Vos, 2020). Sabat et al. (2020) performed a study investigating public sentiment towards restrictive measures in seven European countries. They found that citizens thrusted their government's decisions, and also that people had more worries about the outbreak in southern Europe than in the northern region.

## 2.2 Online shopping

The Internet and communication technologies, in particular telecommuting and online shopping, have been promoted as alternatives allowing people to work and shop from home. Before the pandemic, in 2019, online shopping (also referred to as e-commerce) was estimated to account for about 20.7% of total retail sales worldwide (E-marketer, 2019). In Sweden, as an effect of the COVID-19 pandemic, online shopping turnover was expected to grow by 33% during 2020 compared with 2019 (Gardshol, 2020). Moreover, during the second quarter of 2020 (which corresponds to the period of this study), online shopping volume grew by 49%, and in particular online grocery shopping enjoyed growth of 115% (Gardshol, 2020). In Italy, according to the Osservatorio Politecnico di Milano-Netcomm, online shopping in 2020 will be worth 8% of total retail sales (Digital4, 2020). The overall increase in Italy for 2020 was expected to be 26%, compared to the previous year, while the increase in online grocery shopping during the first wave of the pandemic and the coinciding lockdown was 56% compared to 2019 (Digital4, 2020). In Italy, grocery shops and supermarkets experienced an increase in revenue from online purchases of 300% during the first wave of the pandemic (Lozzi et al., 2020).

Different studies suggest that consumers' online shopping and shopping travel behaviours are significantly influenced by their socio-demographics, experience in using the Internet, car ownership, and geographical factors (Brand et al., 2020; Shi et al., 2019). Shi et al. (2019) found that people who mostly start their shopping trips from inner urban or rural areas were likely to have a higher online shopping share. In line with this, Saphores and Xu (2020) found that households in lower-density areas were less likely to order goods online, compared to their counterparts (however, those who did shop online had a larger number of total deliveries). Whilst online shopping might be seen as a potential solution to mitigate urban congestion, due to the substitution effect of online shopping on shopping trips, it might even increase traffic (Bayarmaa & Susilo, 2014; Lee et al., 2017; Mokhtarian, 2002; Rotem-Mindali & Weltevreden, 2013). Furthermore, previous studies have found that it is people who do not have a private car who are more likely to substitute online shopping for shopping trips (Shi et al., 2019). Brand et al. (2020) explain that shoppers might be attracted to or repelled from online shopping for reasons of convenience, perceived benefits, costs and risks, technology affection, time pressures, and fit into daily schedules (perceived behavioural control), as well as social and environmental dimensions regarding personal norms and beliefs.

Online shopping behaviour changes over time in line with the acceptance of digital technologies (Sunio et al., 2018) and with the change of one's time-use allocations and commitments over time (Schmid, 2019). Saphores and Xu (2020), for example, demonstrate how age and level of education impact the frequency of deliveries from online shopping: younger generations and more educated people order more deliveries. Hoogendoorn-Lanser et al. (2019) found that females shop online more compared to their male counterparts, which is likely due to differing in-home and out-of-home responsibilities and activity engagements.

COVID-19-related restrictions around the world have significantly reduced physical mobility, changed the intra-household arrangement of responsibilities, and subsequently the household members' time-use allocations, which in many ways have shifted behaviour from physical to online shopping alternatives (Hashem, 2020). The pandemic period has also played an important role in improving public acceptance of new technological solutions (e.g., acceptance of autonomous delivery robots for receiving online shopping deliveries) (Pani et al., 2020).

Whilst the behavioural shift to online shopping is clear, evidenced by the increase of online retail sales recorded in the second and third quarters of 2020, to the best of the authors' knowledge, little attention has been given to understanding this shift at the individual level. Therefore, the main scope of the study is to analyse the impacts of the restrictions imposed in the first wave of the pandemic on reported online shopping behaviour, and how individuals from different socio-demographic groups behave differently, given the virus containment strategy in the respondent's country, and the type of purchased goods.

# 3. Methodology

In the following subsections, first, the case study is defined: information is provided on the two countries in terms of restrictive measures, main social differences, and digital maturity. Secondly, the methodology for data collection is described.

## 3.1 Studied areas: Italy and Sweden

Sweden and Italy are two counterparts in Europe in terms of prevention measures. To compare the impacts of the first wave of the pandemic and adaptive behaviours in the two countries, it is important to consider and account for the effects that the policies put in place had on the population. Table 1 shows the different social distancing measures taken by these two countries.

*Table 1 - Restrictive measures in place at the time of data collection (20th April - 18th May 2020). Sources: Ministero della Salute, 2020; Regeringskansliet, 2020.*
*Cited from Bin et al., 2020.*

|  | **Italy** | **Sweden** |
|---|---|---|
| 20th April - 4th May | The country was in lockdown at the release of the survey.<br>• schools and universities were closed;<br>• shops were closed (except grocery shops, pharmacies, tobacconists, newsagents, and petrol stations);<br>• markets were closed;<br>• restaurants, pubs, and cafes were closed (only home delivery allowed);<br>• hair salons, beauticians, and barbers were closed;<br>• banks, post offices, and public offices were open;<br>• parks were closed;<br>• only specific kinds of physical activity outside in constant motion were allowed (e.g., running, jogging);<br>• all factories in non-essential sectors were closed; and<br>• it was forbidden to travel outside the municipality of residence (except for certified working reasons or serious health conditions). | The following restrictions and guidelines were in place at the release of the survey:<br>• all restaurants, bars, cafés, school dining halls, and other venues serving food and beverages had to ensure that tables were spaced appropriately to avoid crowding, and customers had to be always seated while consuming;<br>• it was prohibited to hold public gatherings and public events for more than 50 people;<br>• pharmacies were not allowed to dispense more medications than patients needed for a three-month period;<br>• it was not possible to visit the national care homes for the elderly;<br>• it was forbidden to leave home if experiencing any flu-like symptoms (e.g., coughing, cold, fever);<br>• it was recommended to work from home whenever possible; and<br>• it was strongly recommended to keep a social distance of 2 metres between people when possible. |
| 4th May - 18th May | The country started opening again from 4th May, when the following activities were allowed, while respecting social distancing and using face masks in public places:<br>• to travel in the region of residence for certified reasons (e.g., work, health, visiting relatives);<br>• to travel outside the home region for certified reasons (e.g., work, health, urgent matters, travelling home);<br>• to access parks;<br>• to pick up take-away food; and<br>• to go back to work, for workers in the manufacturing and construction industries, as well as estate agents and wholesalers. | Moreover, from April many companies started a voluntary lockdown of their facilities. |

Moreover, there are differences in both the economic and the social structure of the two countries. Table 2 shows the main social and economic characteristics of Italy and Sweden: population, density, and gender-gap ranking[1] are the most crucial differences, while in terms of employment rate and household structure (average household size by number of people) there is a less significant contrast.

*Table 2 - Economic and population statistics for Italy and Sweden (data from 2019).*

|  | **Italy** | **Sweden** | **Sources** |
| --- | --- | --- | --- |
| *Density* | 200 hab/km$^2$ | 23 hab/km$^2$ | (Worldometer, 2020a, 2020b) |
| *Population* | 60 billion | 10 billion | (Worldometer, 2020a, 2020b) |
| *Gender-gap ranking (worldwide)* | 76th place | 4th place | (World Economic Forum, 2019) |
| *Unemployment rate* | 9.6% (2019) | 6.8% | (Istat, 2020; SCB, 2019) |
| *Average household size* | 2.4 | 2.2 | (Istat, 2020; SCB, 2020) |
| *Life expectancy* | 83.2 | 83.4 | |
| *Average age* | 45.4 | 41.3 | |

According to Harvard Business Review, Sweden was significantly better prepared for remote working in terms of robustness of digital platforms compared to Italy (Chakravorti & Chaturvedi, 2020). According to the study from Chakravorti and Chaturvedi (2020), Sweden is the best positioned among European countries in terms of remote working, due to the robustness of key digital platforms and the use of digital money. On the other hand, Italy is positioned poorly among other European countries, being less resilient and lacking sufficiently robust platforms (Chakravorti & Chaturvedi, 2020). According to Eurostat statistics from 2019, the percentage of people that "usually" worked from home was 3.6% in Italy and 5.9% in Sweden (Eurostat, 2019). An even greater difference can be observed in the percentage of people that "sometimes" worked from home, which is stated to be 1.1% in Italy and 31.3% in Sweden (Eurostat, 2019).

---

[1] Gender-gap ranking is an index defined by the World Economic Forum that allows the tracking of the progress on relative gaps between men and women on health, education, economy and politics (World Economic Forum, 2019).

## 3.2 Survey Description and Methodology

The methodology chosen to address the research questions is an empirical data collection, focusing on behavioural changes before and during the first wave of the pandemic. In this paper, the analysis focuses on behavioural changes connected to shopping activities and explores the differences between Italy and Sweden. Bin et al. (2020) published in a separate study summarised results from an analysis of trade-off behaviours between virtual and physical activities. The responses analysed in this paper were collected between 20th April and 18th May 2020, but include only respondents who live in either Sweden or Italy. To capture the early changes happening during the first wave of the pandemic, the focus was on obtaining quick responses. As there was no choice to be more selective in the recruitment process, the survey was circulated on social media platforms and through the research centre network.

The survey was made available in different languages (including Italian and Swedish), and is divided into six sections: i) changes in travel behaviour relating to daily activities (commuting, grocery shopping, non-grocery shopping, ordering take-away food, eating out, visiting friends and family, going out for entertainment/hobbies, physical activities); ii) changes in internet usage (entertainment, personal calls, work or study, work or study meetings); iii) changes in online shopping behaviour (grocery and non-grocery); iv) perceived safety in performing daily activities (travelling by public transport, travelling by car, visiting shops, being at the workplace or school, going to restaurants, pubs and cafés, going to the gym, spending time outside, receiving home deliveries); v) intention of keeping the new habits (travel and commuting, grocery and non-grocery shopping, work or study, handle meetings at work or school, free time, physical activities) after the pandemic period; and vi) personal information (Bin et al., 2020). In the survey and therefore in the paper, the authors use the term "online shopping" to denote a purchase carried out through online services, with home delivery or delivery at a pickup point (Hoogendoorn-Lanser et al., 2019; Parady et al., 2020). In **Appendix I - Survey structure**, the full content of the survey (i.e., numbered questions) is provided.

To address the objectives mentioned in the introductory section, a series of descriptive analyses were carried out using the software IBM SPSS Statistics 26.0 (IBM Corp., 2019). The following section discusses descriptively: i) the change in the proportion of online and physical shopping behaviours between before and during the first wave of the pandemic, ii) perceived safety during the first wave of the pandemic, and iii) likelihood of keeping the new habits after the pandemic period is over. Then, a series of linear and binary logistic regressions and multinomial logit models were used to measure the impacts of individuals' socio-demographic, built environment, and perceived safety systematically on different types of online shopping behaviours. At first, all the variables are included in the models. The estimation models focus on the significant variables. Therefore, the non-significant variables are systematically removed.

# 4. Description of Survey Results

The total number of answers collected throughout the month of the survey development is 781. However, as this study includes only respondents who live in Italy or Sweden, only 530 responses were used in the analysis. An overview of the demographic distribution of respondents in the two countries is shown in Table 3.

Based on statistical data from SCB for Sweden (SCB, 2020) and Istat for Italy (Istat, 2020), in the survey sample the percentage of males and females differs slightly from the general population. The reached respondents have a higher education level and employment rate. It is important to highlight that the majority of the Italian respondents live in the northern part of the country, where the employment rate is higher: Istat (2020) reports a rate of 66.2% in the given region, which is very close to the employment rate reported by the respondents. The unemployment rate in Sweden is 9.1% according to SCB (2020), and in Stockholm is 6.2% according to the European Commission (European Commission, 2020). Lastly, it can be seen from the dataset that the Italian respondents live in less dense areas compared to their Swedish counterparts. This might seem counterintuitive, since the average density in Italy is 206 hab/km$^2$ (Worldometer, 2020a) and in Sweden it is 25 hab/km$^2$ (Worldometer, 2020b). The reason for this discrepancy relates to how the samples have been collected. In Sweden, the majority of the respondents come from Stockholm and Gothenburg, which are the highest-density areas in the country. Similarly, in Italy, the majority of the respondents come from the northern parts of the country, which is also the highest-density area in the country.

*Table 3 - Overview of the demographic distribution and built environment of the respondents. The statistical data from SCB (SCB, 2020) and Istat (Istat, 2020) is also shown in a dedicated column for both countries.*

| | | Total | Sweden | | | Italy | | |
|---|---|---|---|---|---|---|---|---|
| | | | Survey | | Population Statistics | Survey | | Population Statistics |
| Size of the sample (N) | | 530 (100%) | 212 (40%) | | SCB (2020) | 318 (60%) | | Istat (2020) |
| Gender | Female | 54.3% | 42.5% | | 49.7% | 62.3% | | 51.2% |
| | Male | 45.7% | 57.5% | | 50.3% | 37.7% | | 48.8% |
| Level of education | Less than high school | 2.8% | 0.5% | | 8% | 4.4% | | 38% |
| | High school graduate, diploma or equivalent | 20.9% | 12.7% | | 46% | 26.4% | 95.6% | 62% |
| | Trade/technical/ vocal training | 4.7% | 3.3% | 86.7% | 45% | 5.7% | | |
| | Bachelor's degree | 19.6% | 19.8% | | | 19.5% | | |
| | Master's degree | 40.9% | 41.5% | | | 40.6% | | |
| | Professional degree | 2.5% | 4.2% | | | 1.3% | | |
| | Doctorate degree | 8.5% | 17.9% | | | 2.2% | | |
| Employment | Employee | 56.4% | 62.7% | 72.6% | 67.0% | 52.2% | 67.3% | 57.5% |
| | Self-employed | 10.6% | 7.5% | | | 12.6% | | |
| | Part time worker | 2.5% | 2.4% | | | 2.5% | | |
| | Student | 16.4% | 18.9% | | -- | 14.8% | | -- |
| | Housewife/ houseman | 0.8% | 0.0% | | -- | 1.3% | | -- |
| | Volunteering | 0.6% | 0.0% | | -- | 0.9% | | -- |
| | Military | 0.0% | 0.0% | | -- | 0.0% | | -- |
| | Retired | 10.0% | 7.1% | | -- | 11.9% | | -- |
| | Unemployed | 2.8% | 1.4% | | -- | 3.8% | | -- |
| Density | Average inhabitants per km2 | 2551 | 3637 | | -- | 1827 | | -- |

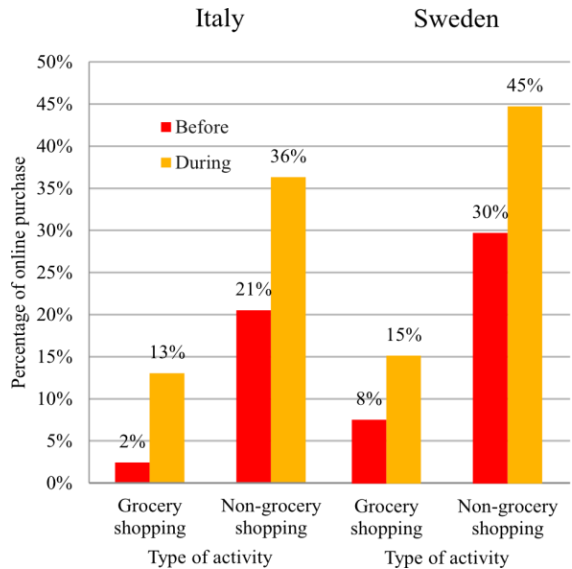

*a) Number of trips[2] per month to grocery and non-grocery shops*

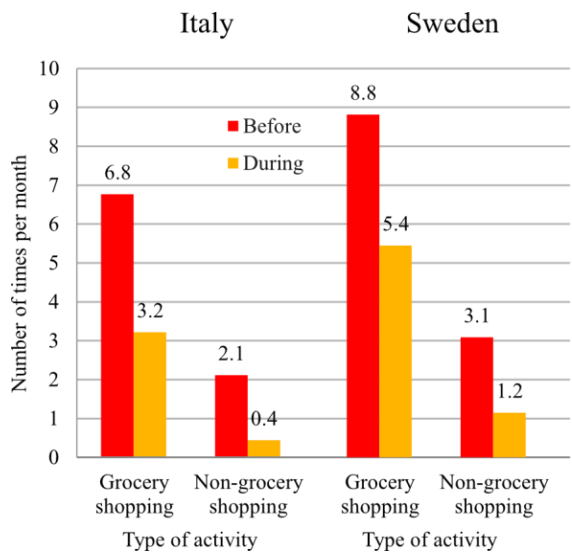

*b) Percentage of online purchase of grocery and non-grocery items*

*Figure 1 - Graphs showing the a) number of trips[2] and b) percentage of online shopping. Numbers are compared before and during the first wave of the pandemic and are divided by country of residence, and represent answers to Q1, Q6, Q10 and Q11 of the survey.*

---

[2] In the paper, *number of trips* refers to trips to a physical shop.

In Figure 1, it can be seen that during the first wave, the number of trips[2] for both grocery and non-grocery shopping decreased and the percentage of online purchases increased among respondents from both Italy and Sweden, compared to before the pandemic. Looking at grocery-related purchases, it is possible to observe a smaller decrease in physical trips (especially in Sweden) compared to non-grocery shopping trips. In Italy, the percentage of online grocery shopping (averaged from all the respondents) increased from 2% to 13%, while in Sweden it was higher than in Italy already before the pandemic, and barely doubled during the first wave, shifting from 8% to 15%. For non-grocery, trips decreased more substantially, especially in Italy. This could be due in part to the stricter policies in place in Italy, and to the fact that only a few non-grocery trips were considered as essential and therefore were carried out despite the risk of visiting the shop. It can also be seen that in Italy, the increase in online shopping was greater than in Sweden.

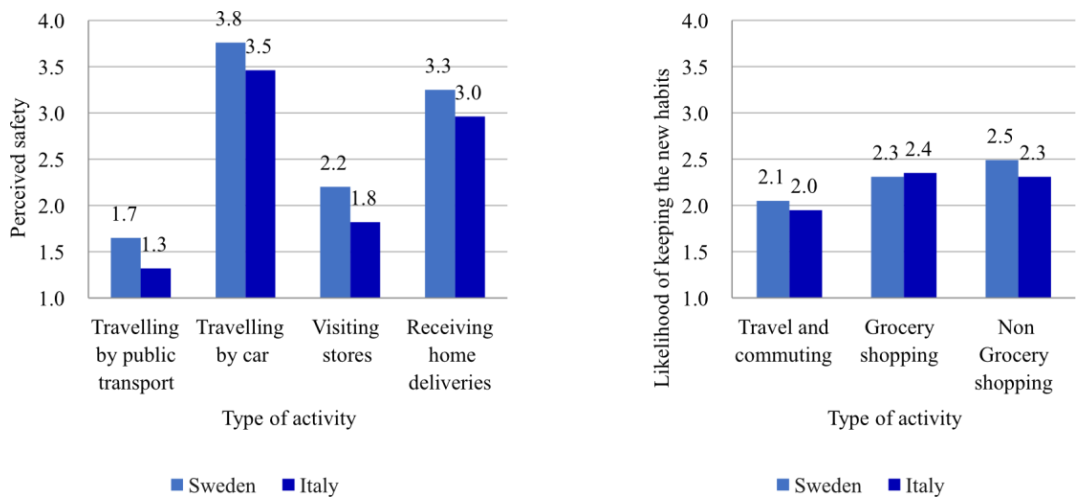

*a) Perceived safety of respondents while performing different activities. In Q13 of the survey, the respondents had the option to choose values between 1 (not safe at all) and 4 (very safe).*

*b) Likelihood of keeping the new habits adopted during the first wave of the pandemic related to different activities. In Q14 of the survey, the respondents had the option to choose values between 1 (not likely) and 4 (very likely).*

*Figure 2 - Graphs of a) perceived safety and b) likelihood of keeping the new habits connected to shopping-related activities, divided by country of residence.*

In Italy, the perceived safety of activities connected to shopping (i.e., travelling by public transport, travelling by car, visiting shops, and receiving home deliveries) is lower than in Sweden (Figure 2a). Moreover, the respondents' answers regarding the likelihood of keeping the new habits connected to shopping (i.e., travel and commuting, grocery and non-grocery shopping) are very similar between the two countries (Figure 2b).

*Table 4 - Purchased non-grocery items during the first wave of the pandemic. In Q12 of the survey, the respondents had the option of selecting multiple types of non-grocery items. The information is divided by country of residence.*

|  | **Sweden** | **Italy** | **Total** |
|---|---|---|---|
| Clothes | 37% | 21% | **27%** |
| Hobby-related items | 58% | 53% | **55%** |
| Products for the house or garden | 47% | 42% | **44%** |
| Work-related items | 16% | 24% | **21%** |

Table 4 shows the percentage of respondents that bought a particular non-grocery item type (i.e., clothes, hobby-related, products for the house or garden, work-related) during the first wave of the pandemic. In Italy, the percentages are slightly lower, especially for clothes. The only type of non-grocery item that was purchased more by the respondents in Italy compared with their Swedish counterparts was work-related. In general, hobby-related items were the most purchased by the respondents.

# 5. Multivariate Analysis

In this section, a series of linear and binary logistic regressions and multinomial logit analyses are presented. First, the percentage of shopping performed online and the items that the respondents reported having bought during the first wave of the pandemic are analysed through linear and binary logistic regressions. In these first models, the dataset has first been analysed as a whole, and then analysed separately between Italian and Swedish respondents, to account for the differences between the two countries presented in Section 3.1. Second, the results of four multinomial logit regressions are presented: two to investigate how the respondents behaved during the first wave of the pandemic in terms of online shopping and shopping trips, and another two to examine how the respondents changed their behaviour compared to before the pandemic. For the second analysis, the authors did not separate Italian and Swedish respondents, as the sample size is too small for the combination of choices.

In Tables 5 and 6, the rows show the independent variables, the columns show the dependent variables that were tested. In Tables 8 and 10, the first column presents the independent variables, while the other columns present the different groups into which the sample is divided. B is the estimated unstandardised regression coefficient and t is the t-test statistics, indicating either the significance of each independent variable to be correlated with the dependent variables (for Tables 5 and 6) or the significance of each variable in influencing belonging to a specific group (for Tables 8 and 10). The variables that are significant at 95% confidence level are indicated by a star (*) while those that are significant at 90% confidence level are indicated by two stars (**). Moreover, in Tables 8 and 10, the positive and significant coefficients indicate the likelihood of a certain demographic characteristic (e.g., Swedish residence) to be significant for belonging to that group, while the negative and significant coefficients indicate the likelihood of being significant for non-belonging to that group.

## 5.1 Online shopping change and non-grocery item selection

Table 5 shows three linear regression models, taking as dependent variables the differences (the value before was subtracted to the one during the first wave of the pandemic) in the percentage of grocery and non-grocery shopping that is performed online. The first model refers to the dataset as a whole, while the second and third models refer to the split datasets of Italian residents and Swedish residents, respectively. It can be seen that there are significant differences between the two countries. For example, in the Italian dataset, being highly educated is correlated (at 90% confidence level) with a greater increase in non-grocery online shopping, compared to their Swedish counterparts. In Italy, being a worker is correlated with a greater increase in online non-grocery shopping (coef. 0.137), while in Sweden, being a worker or student is correlated with a lower increase in online grocery shopping (coef. -0.272 and -0.274, respectively). Being a student in Sweden is correlated with a lower increase in online non-grocery shopping (coef. -0.196). For grocery items, it can be seen that only for the Swedish dataset a greater decrease in physical trips is correlated with a greater increase in online shopping. For non-grocery items, however, in Italy, the decrease in physical shopping trips is associated (at 90% confidence level) with an increase in online shopping for both groceries and non-groceries. It can also be seen from the table that people who felt safer in physical shops are correlated with a greater increase in online purchases when compared to their counterparts; in Italy this significance is valid for both grocery and non-grocery items, while in Sweden it is valid for grocery items. The household structure and the perceived safety of travelling either by public transport or by car, and of receiving home deliveries, does not seem to be significant in this analysis. When considering the population density in the place of residence, it can be seen that in Italy, people who live in more densely-populated areas are correlated with a greater increase in online shopping compared to their counterparts. This correlation was not found in the Swedish dataset.

*Table 5 - Linear regression on the difference in percentage of online shopping (for both grocery and non-grocery), comparing before and during the first wave of the pandemic.*

| | | Grocery | | | Non-grocery | | |
| --- | --- | --- | --- | --- | --- | --- | --- |
| | | All | Italy | Sweden | All | Italy | Sweden |
| | Ranges | Coef. | Coef. | Coef. | Coef. | Coef. | Coef. |
| **(Constant)** | - | 0.131 | 0.057 | 0.320 | 0.046 | 0.018 | 0.146 |
| **Worker** | [0,1] | -0.055** | 0.032 | -0.272* | 0.093* | 0.137* | -0.036 |
| **Student** | [0,1] | -0.081* | -0.002 | -0.274* | -0.050 | 0.005 | -0.196* |
| **Highly educated** | [0,1] | n.s. | n.s. | n.s. | 0.046** | 0.059** | 0.018 |
| **Change in grocery trips** | [-100%, +100%] | -0.002 | 0.002 | -0.006* | 0.004** | 0.004 | 0.005 |
| **Change in non-grocery trips** | [-100%, +100%] | -0.005 | -0.009** | -0.003 | -0.014* | -0.011** | -0.019* |
| **Feel safe in shops** | [0,1] | -0.082* | -0.064* | -0.081* | -0.054* | -0.088* | 0.006 |
| **Population density in place of residence** | [0,+∞] | 8.0E-5* | 1.9E-4* | 1.7E-5 | 8.6E-5** | 1.3E-4* | 8.0E-5 |
| | | | | | | | |
| **Adjusted R Square** | - | 0.035 | 0.044 | 0.135 | 0.075 | 0.091 | 0.080 |
| **Std. Error of the Estimate** | - | 0.227 | 0.214 | 0.230 | 0.259 | 0.263 | 0.251 |

*Legend. The more positive the coefficient, the larger the relative increase in online purchase, compared to their counterparts. One star (\*) indicates 95% confidence level, two stars (\*\*) indicates 90% confidence level. Variable ranges are indicated for each variable. "n.s." indicates that the variable is non-significant and is not included in the model.*

In Table 6, it can be seen that females in Sweden were more likely to buy clothes during the first wave of the pandemic than their male counterparts (coef. 0.850); at the same time, males in Sweden were more likely to buy hobby-related items (coef. -1.146). In both countries, feeling safe on public transport and travelling by car is correlated with buying clothes (coef. 0.743 and 1.042 respectively). Being a student or worker in Italy is correlated with the choice to buy work-related items (coef. 1.764 and 1.912 respectively), while it is correlated with the choice not to buy house- or garden-related items compared to their counterparts (coef. -0.597 and -1.637 respectively). In both countries, having elderly members in the households is correlated with buying more work-related items (coef. 0.755). Moreover, in both countries, respondents who reported feeling safe in shops are also found to have a higher probability of buying work-related items (at 90% confidence level, coef. 0.436). Among Swedish respondents, the population density (calculated with respect to where the respondents live) was found significant (at 90% confidence level): living in a denser area is marginally correlated with buying hobby-related items and not buying items for the house or garden.

*Table 6 - Binary logistic regression of the variables related to which types of items the respondents purchased during the first wave of the pandemic.*

| | Ranges | Clothes | | | Hobby | | | House | | | Work | | |
|---|---|---|---|---|---|---|---|---|---|---|---|---|---|
| | | All | Italy | Sweden | All | Italy | Sweden | All | Italy | Sweden | All | Italy | Sweden |
| | | Coef. | Coef. | Coef. | Coef. | Coef. | Coef. | Coef. | Coef. | Coef. | Coef. | Coef. | Coef. |
| (Constant) | - | -2.817 | -2.613 | -2.799 | 0.317 | -0.043 | 5.501 | -0.076 | 0.511 | -1.022 | -3.389 | -3.311 | -4.561 |
| Female | [0,1] | 0.312 | 0.104 | 0.850* | -0.638* | -0.340 | -1.146* | n.s. | n.s. | n.s. | n.s. | n.s. | n.s. |
| Worker | [0,1] | n.s. | n.s. | n.s. | 0.166 | 0.514 | -2.581* | -0.348 | -0.597** | 0.211 | 1.189* | 1.764* | 0.703 |
| Student | [0,1] | n.s. | n.s. | n.s. | -0.056 | 0.539 | -3.519* | -1.342* | -1.637* | -0.761 | 1.322* | 1.912* | 1.031 |
| Highly educated | [0,1] | n.s. | n.s. | n.s. | 0.072 | 0.304 | -0.796** | n.s. | n.s. | n.s. | n.s. | n.s. | n.s. |
| Have children in the household | [0,1] | 0.530* | 0.558** | 0.201 | n.s. | n.s. | n.s. | n.s. | n.s. | n.s. | n.s. | n.s. | n.s. |
| Have adults in the household | [0,1] | n.s. | n.s. | n.s. | n.s. | n.s. | n.s. | 0.175 | -0.132 | 0.670* | n.s. | n.s. | n.s. |
| Have elderly in the household | [0,1] | n.s. | n.s. | n.s. | -0.348 | -0.523** | -2.021 | n.s. | n.s. | n.s. | 0.755* | 0.407 | 1.786** |
| Feel safe in stores | [0,1] | -0.460* | -0.738** | -0.363 | n.s. | n.s. | n.s. | n.s. | n.s. | n.s. | 0.436** | 0.551** | 0.665** |
| Feel safe to receive home deliveries | [0,1] | n.s. | n.s. | n.s. | 0.599* | 0.532* | 0.596 | n.s. | n.s. | n.s. | n.s. | n.s. | n.s. |
| Feel safe to travel with public transport | [0,1] | 0.743* | 1.325* | 0.124 | -0.568** | -0.090 | -0.666 | n.s. | n.s. | n.s. | n.s. | n.s. | n.s. |
| Feel safe to travel with car | [0,1] | 1.042* | 0.611 | 1.733 | n.s. | n.s. | n.s. | n.s. | n.s. | n.s. | n.s. | n.s. | n.s. |
| Density of population in place of residence | [0,+∞] | n.s. | n.s. | n.s. | 4.0E-4 | 1.2E-4 | 0.001** | -1.6E-4 | 2.1E-4 | -0.001** | n.s. | n.s. | n.s. |
| | | | | | | | | | | | | | |
| -2 Log likelihood | - | 600.892 | 317.351 | 263.691 | 700.232 | 421.413 | 256.795 | 707.125 | 417.668 | 276.995 | 528.763 | 334.042 | 179.990 |
| Cox & Snell R Square | - | 0.039 | 0.031 | 0.069 | 0.052 | 0.055 | 0.136 | 0.038 | 0.047 | 0.074 | 0.028 | 0.055 | 0.031 |
| Nargel kerke R Square | - | 0.057 | 0.049 | 0.095 | 0.069 | 0.073 | 0.183 | 0.050 | 0.063 | 0.098 | 0.044 | 0.082 | 0.053 |
| Chi-Square | | 21.108 | 10.108 | 15.234 | 28.144 | 17.906 | 30.956 | 20.342 | 15.280 | 16.220 | 15.235 | 18.005 | 6.695 |

*Legend. The more positive the coefficient, the larger the likelihood of buying the type of item, compared to their counterparts. One star (\*) indicates 95% confidence level, two stars (\*\*) indicates 90% confidence level. Variable ranges are indicated for each variable. "n.s." indicates that the variable is non-significant and is not included in the model.*

## 5.2 Respondents grouping based on shopping behaviour

To address the research questions and to analyse the phenomena of social exclusion and behavioural change systematically, the combined sample (both Swedish and Italian responses) was divided into different groups according to two grouping methodologies, i.e.: one based on the absolute number of trips made and the percentage of online shopping done during the first wave of the pandemic, and the other based on the differences in the respondents' reactions (in terms of the number of physical trips and the difference in online shopping) between during and before the first wave. The next two subsections explore these two grouping methodologies.

### 5.2.1 Activities during the first wave of the pandemic

Table 7 presents the different groups into which the sample was divided for the analysis. Group 0 is the reference group for the following multinomial logit regression analysis, and it includes respondents who did not travel (physically) for shopping purposes[3] and did not buy online during the first lockdown period. Looking at the demographic of this group, it can be seen that more than 80% of the respondents that fall into this group are Italian. Group 1 represents respondents who continued to visit shops but did not use any e-commerce services. Group 2 includes respondents who did not travel to visit shops but did make purchases online. Finally, Group 3 represents people who continued to visit shops and also used e-commerce services.

*Table 7 - Grouping methodology for a) grocery shopping and b) non-grocery shopping. The grouping methodology in this case is related to behaviours of the respondents during the first wave of the pandemic. The number of respondents per group is shown in brackets.*

*a) Grocery shopping*

|  | Online during=0 | Online during>0 |
|---|---|---|
| # trips =0 | Group 0 (48) | Group 2 (25) |
| # trips >0 | Group 1 (309) | Group 3 (148) |

*b) Non-grocery shopping*

|  | Online during=0 | Online during>0 |
|---|---|---|
| # trips =0 | Group 0 (90) | Group 2 (259) |
| # trips >0 | Group 1 (31) | Group 3 (150) |

---

[3] The frequency option for the number of trips "less than once a month" is here approximated to 0.

The results of the multinomial logit model shown in Table 8a for grocery shopping show that female respondents are more likely to be part of Groups 1 or 3, which means they are more likely to visit grocery shops. Moreover, living in Sweden is correlated with a higher likelihood of belonging to Groups 1, 2 or 3 (at 90% confidence level), which means respondents are still shopping either online or physically or both. Finally, students are unlikely to be part of Group 3 and 1 (at 90% confidence level), people who felt safe in shops are likely to be in Group 1 (at 90% confidence level).

The results of the multinomial logit model presented in Table 8b for non-grocery shopping show that having children in the household is correlated with being part of Groups 1, 2 and 3. Belonging to Group 1 and 3 is correlated with living in Sweden. Group 1 is also negatively correlated with having elderly in the household, which is probably because the respondents fear infecting their older household members. Belonging to Group 1 is also negatively correlated with being a worker and a student. Being part of Group 2 is correlated with being a worker and with feeling safe receiving home deliveries. Finally, belonging to Group 3 is correlated with being male, a worker, and feeling safe receiving home deliveries.

*Table 8 - Multinomial logit regressions for a) grocery shopping and b) non-grocery shopping, according to the groups in Table 7. The reference category is group 0.*

*a) Grocery shopping*

|  | Range | Group 1 B | Group 2 B | Group 3 B |
|---|---|---|---|---|
| **Intercept** | -- | 1.086 | -1.438 | 0.714 |
| **Female** | 0, 1 | 0.660* | 0.497 | 0.718* |
| **Swedish residence** | 0, 1 | 1.417* | 2.102* | 0.815** |
| **Student** | 0, 1 | -0.668** | -0.400 | -1.409* |
| **Feel safe in shops** | 0, 1 | 0.784** | -1.632 | 0.497 |

|  | -2 Log Likelihood | Chi-Square | df | Sig. |
|---|---|---|---|---|
| **Intercept Only** | 181.239 |  |  |  |
| **Final** | 124.825 | 56.414 | 12 | <.001 |

|  | Pseudo R-Square |
|---|---|
| **Cox and Snell** | .101 |
| **Nagelkerke** | .116 |
| **McFadden** | .052 |

*b) Non-grocery shopping*

|  | Range | Group 1 B | Group 2 B | Group 3 B |
|---|---|---|---|---|
| **Intercept** | -- | -0.581 | -0.473 | -2.599 |
| **Density** | [0, ∞] | 0.002* | 0.001 | 4.8E-4 |
| **Female** | 0, 1 | -0.468 | -0.064 | -0.650* |
| **Have children in the household** | 0, 1 | 1.366* | 1.079* | 1.014* |
| **Have adults in the household** | 0, 1 | -0.156 | 0.252 | 0.896* |
| **Have elderly in the household** | 0, 1 | -1.908* | -0.055 | -0.049 |
| **Swedish residence** | 0, 1 | 1.327* | -0.074 | 1.647* |
| **Worker** | 0, 1 | -1.441* | 0.885* | 1.069* |
| **Student** | 0, 1 | -1.745* | 0.591 | -0.310 |
| **Feel safe to receive home deliveries** | 0, 1 | -0.265 | 0.548* | 1.222* |

|  | -2 Log Likelihood | Chi-Square | df | Sig. |
|---|---|---|---|---|
| **Intercept Only** | 1089.326 |  |  |  |
| **Final** | 915.946 | 173.379 | 27 | <.001 |

|  | Pseudo R-Square |
|---|---|
| **Cox and Snell** | .279 |
| **Nagelkerke** | .308 |
| **McFadden** | .139 |

*Legend. The more positive the coefficient, the larger the likelihood of belonging to the group, compared to their counterparts. One star (*) indicates 95% confidence level, two stars (**) indicates 90% confidence level. Variable ranges are indicated for each variable.*

### 5.2.2 Difference in behaviour between during and before the first wave of the pandemic

This subsection focuses on behavioural change: who increased or decreased their number of shopping trips and who increased or decreased their online shopping activity. In the previous subsection, the respondents were grouped based on their number of shopping trips and online shopping percentage during the first wave of the pandemic, whilst in this subsection, the respondents are grouped based on changes in behaviour during the first wave compared to the period before the pandemic.

*Table 9 - Grouping methodology for a) grocery shopping and b) non-grocery shopping. The grouping methodology in this case is related to the difference in behaviour between during and before the first wave of the pandemic. The number of respondents per group is shown in brackets.*

*a) Grocery shopping*

|  | Online difference<0 | Online difference=0 | Online difference>0 |
|---|---|---|---|
| # trips difference<0 |  | Group 1 (195) | Group 3 (81) |
| # trips difference=0 | Group 0 (23) | Group 2 (148) | Group 4 (50) |
| # trips difference>0 |  | Group 5 (33) | |

*b) Non-grocery shopping*

|  | Online difference<0 | Online difference=0 | Online difference>0 |
|---|---|---|---|
| # trips difference<0 |  | Group 1 (172) | Group 3 (164) |
| # trips difference=0 | Group 0 (35) | Group 2 (84) | Group 4 (49) |
| # trips difference>0 |  | Group 5 (26) | |

Table 9 shows the different groups into which the respondents have been divided for this second part of the analysis. Group 0, which represents respondents who reduced their online shopping activity regardless of the difference in the number of physical trips, is relatively small (23 people for grocery and 35 people for non-grocery shopping) and represents the reference group in the multinomial logit estimation. Group 2 represents the respondents who showed minimal change in both types of behaviour[4], and is one of the larger groups when considering grocery shopping. Groups 1 and 4 represent, respectively, the respondents who showed minimal change in one of the two types of behaviours, but either decreased their number of trips or increased their online shopping. Group 1 is the largest for both grocery and non-grocery shopping. Group 3 represents the respondents who increased their online shopping and at the same time decreased their number of trips. Lastly, Group 5 represents the respondents who showed an increase in the difference in the number of physical trips, and either had minimal change in their online shopping behaviour or increased their online shopping during the first wave of the pandemic; this group is also relatively small. The decision to combine all the respondents with a decrease in online shopping for Group 0 and with an increase in the number of trips for Group 5 is motivated by the fact that otherwise groups with too few respondents would have been created.

---

[4] The trip difference is 0 when the respondents chose the same frequency option for both before and during the pandemic. However, this does not necessarily mean that they have not changed their habits, but rather that they had a minimal change in habits.

The results of the multinomial logit estimation shown in Table 10a) for grocery shopping show that respondents without children in the household are more likely to belong, in order of relevance, to Group 5, Group 2, Group 1 and Group 3. Moreover, respondents without elderly in the household are more likely to belong to Group 2. Swedish respondents and highly educated people (at 90% confidence level) are more likely to belong to Group 4.

The results of the multinomial logit estimation shown in Table 10b) for non-grocery shopping show that workers are more likely to belong, in order of relevance, to Group 3, Group 2, Group 1 and Group 4. Being highly educated is negatively correlated with Groups 1, 2 and 5 (at 90% confidence level). Moreover, people who feel safe going to shops are more likely to be part of Group 2. These people probably had a minimal change in their behaviour because they felt safe to continue doing part of their non-grocery shopping physically in shops.

*Table 10 - Multinomial logit regressions for a) grocery shopping and b) non-grocery shopping, according to the groups in Table 9. The reference category is group 0.*

*a) Grocery shopping*

|  |  | Group 1 | Group 2 | Group 3 | Group 4 | Group 5 |
|---|---|---|---|---|---|---|
|  | Ranges | B | B | B | B | B |
| **Intercept** | -- | 3.515 | 3.321 | 2.394 | 2.445 | 2.127 |
| **Have children in the household** | 0, 1 | -1.191* | -1.186* | -1.306* | -0.553 | -1.321* |
| **Have elderly in the household** | 0, 1 | -0.429 | -1.651* | -0.391 | 0.063 | -2.233** |
| **Swedish residence** | 0, 1 | -0.526 | 0.186 | -0.507 | -1.433* | -0.487 |
| **Highly educated** | 0, 1 | -0.662 | -1.068 | -0.315 | -1.210** | -0.981 |

|  | -2 Log Likelihood | Chi-Square | df | Sig. |
|---|---|---|---|---|
| **Intercept Only** | 263.317 |  |  |  |
| **Final** | 200.792 | 62.525 | 20 | <.001 |

| Pseudo R-Square | |
|---|---|
| **Cox and Snell** | .111 |
| **Nagelkerke** | .117 |
| **McFadden** | .038 |

*b) Non-grocery shopping*

| | Ranges | **Group 1** B | **Group 2** B | **Group 3** B | **Group 4** B | **Group 5** B |
|---|---|---|---|---|---|---|
| **Intercept** | -- | 2.238 | 0.921 | 1.135 | 0.415 | -0.376 |
| **Have adults in the household** | 0, 1 | -1.056** | -0.942 | -0.795 | -0.703 | -0.232 |
| **Swedish residence** | 0, 1 | 0.466 | 0.869** | 0.420 | 0.276 | 0.989** |
| **Worker** | 0, 1 | 1.268* | 1.495* | 1.889* | 1.459* | 1.011** |
| **Highly educated** | 0, 1 | -1.010* | -0.990* | -0.347 | -0.563 | -1.059** |
| **Feel safe in shops** | 0, 1 | 0.687 | 1.215* | 0.382 | 0.275 | 0.692 |

| | -2 Log Likelihood | Chi-Square | df | Sig. |
|---|---|---|---|---|
| **Intercept Only** | 407.773 | | | |
| **Final** | 349.269 | 58.504 | 25 | <.001 |

| Pseudo R-Square | |
|---|---|
| **Cox and Snell** | .105 |
| **Nagelkerke** | .109 |
| **McFadden** | .035 |

*Legend. The more positive the coefficient, the larger the likelihood of belonging to the group, compared to their counterparts. One star (\*) indicates 95% confidence level, two stars (\*\*) indicates 90% confidence level. Variable ranges are indicated for each variable.*

# 6. Discussion

The overall results of the survey point to an increase in online shopping and a decrease in the number of physical shopping trips, as expected and in accordance with the literature presented (Digital4, 2020; Gardshol, 2020; Hashem, 2020). In line with the stricter restrictions in Italy, relatively compared to Sweden (Table 1), Italian respondents reduced their number of trips more compared to their Swedish counterparts (Figure 1). An increase in online shopping was also observed for both countries, as expected from the literature (Digital4, 2020; Gardshol, 2020); in Italy the results show a greater increase during the first wave of the pandemic compared to the period before the pandemic. Two factors play a role in the increase in online shopping: i) in Sweden there were less stringent restrictions in place (Table 1), and residents were always allowed to visit physical shops; and ii) the already higher (compared to Italy) figures in online shopping purchases (Digital4, 2020; Gardshol, 2020).

Moreover, in both Italy and Sweden, respondents tended to reduce their non-grocery shopping trips due to doing more online shopping, and this supports the claim that online shopping has a substitution effect on the frequency of shopping trips (Mokhtarian, 1990; Shi et al., 2019). However, in Italy the correlation is less strong; this may be due to a combination of more stringent restrictions and a lower level of technology readiness, but also to the more critical economic crisis in Italy compared with Sweden (Eurostat, 2020).

**Answers to research questions**

Overall, respondents who indicated that they felt unsafe in shops were more likely to shop online (Table 5). In Italy, the correlation is more significant for non-grocery shopping, and in Sweden for grocery shopping. This could be discussed in terms of the policies in place in the two countries: in Italy, non-grocery shops were closed (Table 5, Ministero della Salute, 2020).

In Italy, highly educated people and workers are shown to have increased their non-grocery online shopping activity (Table 5 and Table 8b), which is in line with findings from Hashem (2020) and Hoogendoorn-Lanser et al. (2019). In Sweden, an opposite trend can be seen: both workers and students showed a smaller increase in their grocery online shopping activity compared to their counterparts (Table 5). Being a worker is correlated with more transitions towards buying more non-grocery items online, as shown in Table 8b, where it can be seen that workers mainly belong to Groups 2 and 3 (and therefore are actively shopping online).

In Italy, the respondents bought more work-related items compared to their Swedish counterparts (Table 4): as Harvard Business Review points out, the population was not prepared for remote working (Chakravorti & Chaturvedi, 2020), and therefore had a need to purchase items related to work. Another related result is that in Italy, being either a student or a worker is correlated with the choice to buy work-related items. An important statistic that is worth mentioning is the percentage of people that worked from home before the pandemic started, which according to Eurostat amounted to 37.2% in Sweden and only 4.7% in Italy (Eurostat, 2019). When the lockdown period started in Italy, in spring 2020, a new regulatory framework for remote working was implemented (Ministero del Lavoro e delle Politiche Sociali, 2020). In Italy, there was hesitation towards remote working, associated with a resistance to allowing employees to work independently, without the constant supervision that there might be in a regular office space (Betti, 2020).

Females are more likely to have kept the habit of going to the shops for groceries, as it appears in the multinomial logit model (Table 8a). Unlike the result shown in Hoogendorn-Lanser et al. (2019), Hashem (2020) and Saphores and Xu (2020), the analysis does not show that females are associated with more online shopping (as it can be seen in Table 5, female is removed as a variable since it is non-significant), neither for groceries nor for non-groceries.

The household structure is not relevant in the first analysis regarding online shopping, but it is relevant in the second one, related to the kinds of bought items. Those households with elderly members are shown to have a correlation with the choice to buy work-related items; this could be a sign that, given that the pandemic has had a greater effect on older people (CDC, 2020), the people around them were taking the necessary precautions to socially distance themselves. The households with children are shown to have a correlation with the choice to buy clothes, compared to their counterpart. Moreover, from Table 8b, households without children were more likely to have stopped going to shops compared to their counterparts. The fact that households with children still shopped for non-grocery items could be a sign that they regarded non-grocery shopping as essential. This is also in line with the National Retail Federation consumer view, which confirms that children influence their parents' shopping behaviours. The two top categories of influence are toys and clothes (NRF, 2019).

When looking deeper into behaviours during the first wave of the pandemic, it can be seen that the group of people who stopped both types of grocery shopping activity (online and physical) is extremely small (Table 7a). Demographic analysis of the group reveals that most of the respondents are Italian residents. These people (i.e., elderly, children, young adults) did not buy groceries either online or in physical shops, therefore they might either have needed external help or have had other members of the household in charge of buying essential items for them. It can also be seen that respondents who felt safe in shops are more likely to belong to Group 1 (at 90% confidence level), where the only way of shopping is to go to the physical shop. For non-grocery shopping, the group of people who stopped both types of activity is double the size (Table 7b) of the grocery shopping one (Table 7a). This could simply be related to different people's perceptions of which types of shopping are essential. Being a Swedish resident is correlated with being part of Groups 1 and 3 (Table 8b), who still chose to go to physical shops. This is in line with the considerations on policies in Sweden (Table 1). Moreover, feeling safe receiving home deliveries is a determining factor in the decision of the respondents to shop online.

**Policy reflections**

The importance of technological readiness in coping with disruptive events has been highlighted by the COVID-19 pandemic: it has been seen in this study to facilitate the transition from physical activities to online alternatives. However, whilst there have been behavioural changes towards online shopping, such changes cannot be taken as granted. If stakeholders wish to maintain these patterns, support systems and policies that promote these behaviours need to be established. The results of this study also highlight the importance of conveying a clear and consistent message about which behaviours are safe, in order for people to have more confidence in the opportunities provided.

As household structure has been seen to influence behavioural change (for example, households with children are more likely to continue going to physical shops), if the government wishes to change travel behaviour among such a socio-demographic group, a tailored support system should be provided. On intra-household arrangement of responsibilities, it can be seen that there was a redistribution of intra-household chores, in particular shopping. In the past, shopping was done mainly by females, but now there could be a change in intra-household values, leading to a fairer distribution of chores within the household. This is shown in Chung et al. (2021), who reported a rise in the number of fathers working from home, which may have shifted towards a fairer distribution of chores. Moreover, in Del Boca et al. (2020), men whose partners worked physically at their usual workplace spent more time on housework during the COVID-19 pandemic compared to before.

**Caveats**

As discussed in Section 4, the survey sample has a higher education level and a lower unemployment rate than the corresponding averaged national statistics. This, which might be due to the way the respondents were recruited, results in a bias since the sample does not represent the national population. Most of the Italian respondents come from northern Italy and most of the Swedish respondents come from either Stockholm or Gothenburg. Moreover, the survey was conducted online. This might have excluded individuals who do not engage in online activities. Therefore, the results should be interpreted with caution. In the analysis, the authors did not distinguish between sub-periods correlated with different policies in place, which could have an effect on the respondents' answers.

Population density is significant only in Italy for online shopping activities; participants living in denser areas are more engaged in online shopping activities compared to their counterparts (Table 5). In all other analyses, population density was found not significant, which is unique compared to previous evidence on the impacts of geographical location on the propensity for doing online shopping (Saphores & Xu, 2020; Shi et al., 2019). However, the fact that for the remaining analysis density is not relevant could be biased by the spread of the sample among urban and rural areas: the average density for both Swedish and Italian respondents is higher than the average density in Sweden and Italy respectively, and in Sweden the variance of the density is low, with the majority sample from high density areas. It could be also due to the nature of the pandemic-driven lockdown being deployed, or the size of the sample.

# 7. Conclusion

Based on 530 responses from an online survey, this study investigates the impacts of the first wave of the COVID-19 pandemic on the behavioural transition from physical to online shopping. The descriptive results and multivariate analyses have been used to explore who changed their shopping behaviour the most, how respondents adopted different strategies, what the main differences are between Italy and Sweden, and the influence of population density on behavioural change. The study suggests that in Italy, there has been a larger increase in online shopping and a bigger reduction in physical trips compared to Sweden, which is in line with the reflections at the beginning of the paper on differences in policies and online shopping behaviours prior to the pandemic.

The main takeaways are the following:

1. In Italy, the respondents bought work-related items more compared to their Swedish counterparts. It is also shown in the results that in Italy, workers and students were more likely to make the choice to buy work-related items. This is presumably because, prior to the pandemic period, many Italian employees were not prepared for remote working and therefore many respondents were in need of such items. Different policies were implemented in Italy during the first wave of the pandemic to help workers adapt to the new working situation.

2. The analysis shows that the respondents were impacted differently in terms of trips. Some respondents may have unconsciously excluded themselves from doing certain activities, since they reported not doing any grocery nor non-grocery shopping during the first wave of the pandemic. The results also show that the group of respondents who did not shop for non-grocery items is larger than the group that did not shop for grocery items, which highlights the discrepancies between which types of shopping the respondents may consider as essential.

3. The household structure of the respondents influenced their behaviour in different ways. Having elderly members in the household impacted the precautions that the other household members took, including being prepared to work remotely and avoiding going to physical shops to buy non-grocery items. In this sense, the households with elderly members can also be categorised as a risk group who may end up being socially excluded – a group among the population which has not been the focus of the current support policies. On the other hand, having children in the household is correlated with continuing to visit non-grocery physical shops.

In summary, the results highlight the differences between the two study areas (i.e., Sweden and Italy) in terms of social distancing measures, social structures, and technology readiness. These confirm the need for different policy strategies to help different groups of citizens to comply with pandemic-related measures, whilst still being able to fulfil their individual shopping needs (which are highly influenced by their socio-demographic characteristics, technology readiness, and household composition). The study also shows the presence of discrepancies between what one may consider as essential activities. All of these findings highlight that different socio-demographic and household structures matter in determining the direction of shopping behaviour transformation during the COVID-19 pandemic.

# Declarations

### Availability of data and materials

The datasets collected and analysed during the current study are not yet publicly available but are available from the corresponding author upon request.

### Competing interests

The authors declare that they have no competing interests.

### Funding

This research is supported by ITRL (Integrated Transport Research Lab, KTH Royal Institute of Technology, Sweden) and DAVeMoS (the FFG/BMK Endowed Professorship programme, grant number 862678, at the University of Natural Resources and Life Sciences – BOKU, Vienna).

### Authors' contributions

All authors contributed to the ideation of the study.

CA and EB designed the survey under supervision of AP and YS.

CA, EB and YS analysed the data and interpreted the results.

All authors reviewed existing literature.

All authors contributed to the writing phase and approved the final manuscript.

**Acknowledgements**

The authors would like to thank the volunteers who helped in translating, reviewing and distributing the survey.

# References

Almlöf, E., Rubensson, I., Cebecauer, M., & Jenelius, E. (2020). *Who is still travelling by public transport during COVID-19? Socioeconomic factors explaining travel behaviour in Stockholm based on smart-card data.* https://doi.org/10.13140/RG.2.2.26330.36805

Apple. (2020). *COVID-19 – Mobility Trends Reports.* Apple. https://www.apple.com/covid19/mobility

Arimura, M., Ha, T. V., Okumura, K., & Asada, T. (2020). Changes in urban mobility in Sapporo city, Japan due to the Covid-19 emergency declarations. *Transportation Research Interdisciplinary Perspectives*, *7*, 100212. https://doi.org/10.1016/j.trip.2020.100212

Bayarmaa, A., & Susilo, Y. O. (2014). Telecommuting and telecommunications. In M. Garrett (Ed.), *Encyclopedia of Transportation: Social Science and Policy* (Vol. 1, pp. 1320–1322). Thousand Oaks, CA: SAGE Publications. https://sk.sagepub.com/reference/encyclopedia-of-transportation/n462.xml

Beck, M. J., & Hensher, D. A. (2020). Insights into the impact of COVID-19 on household travel and activities in Australia – The early days of easing restrictions. *Transport Policy*, *99*,

95–119. https://doi.org/10.1016/j.tranpol.2020.08.004

Betti, I. (2020, February 15). *Coronavirus impone maxi-test mondiale sullo smart working. De Masi: 'In Italia c'è una resistenza patologica'.* https://www.huffingtonpost.it/entry/coronavirus-impone-maxi-test-mondiale-sullo-smart-working-de-masi-in-italia-ce-una-resistenza-patologica_it_5e440ac7c5b61b84d3433541

Bin, E., Andruetto, C., Susilo, Y., & Pernestål, A. (2020). The Trade-Off Behaviours between Virtual and Physical Activities during COVID-19 Pandemic Period. *SSRN Electronic Journal.* https://doi.org/10.2139/ssrn.3698595

Brand, C., Schwanen, T., & Anable, J. (2020). 'Online Omnivores' or 'Willing but struggling'? Identifying online grocery shopping behavior segments using attitude theory. *Journal of Retailing and Consumer Services*, *57*, 102195. https://doi.org/10.1016/j.jretconser.2020.102195

CDC. (2020, February 11). *Coronavirus Disease 2019 (COVID-19).* Centers for Disease Control and Prevention. https://www.cdc.gov/coronavirus/2019-ncov/need-extra-precautions/older-adults.html

Chakravorti, B., & Chaturvedi, R. S. (2020, April 29). *Which Countries Were (And Weren't) Ready for Remote Work?* https://hbr.org/2020/04/which-countries-were-and-werent-ready-for-remote-work

Chan, H. F., Moon, J. W., Savage, D. A., Skali, A., Torgler, B., & Whyte, S. (2020). *Can psychological traits explain mobility behavior during the COVID-19 pandemic?* [Preprint]. PsyArXiv. https://doi.org/10.31234/osf.io/5q3jv

Chung, H., Birkett, H., Forbes, S., & Seo, H. (2021). Covid-19, Flexible Working, and Implications for Gender Equality in the United Kingdom. *Gender & Society*, *35*(2), 218–232. https://doi.org/10.1177/08912432211001304

Dahlberg, M., Edin, P.-A., Grönqvist, E., Lyhagen, J., Östh, J., Siretskiy, A., & Toger, M. (2020). *Effects of the COVID-19 Pandemic on Population Mobility under Mild Policies: Causal Evidence from Sweden*. 32.

De Vos, J. (2020). The effect of COVID-19 and subsequent social distancing on travel behavior. *Transportation Research Interdisciplinary Perspectives*, *5*, 100121. https://doi.org/10.1016/j.trip.2020.100121

Del Boca, D., Oggero, N., Profeta, P., & Rossi, M. (2020). Women's and men's work, housework and childcare, before and during COVID-19. *Review of Economics of the Household*, *18*(4), 1001–1017. https://doi.org/10.1007/s11150-020-09502-1

Digital4. (2020, July 8). *Acquisti online, crescita record grazie al lockdown: +26%*. Digital4. https://www.digital4.biz/marketing/ecommerce/acquisti-online-2020-polimi-netcomm/

E-marketer. (2019). *Global Ecommerce 2019*. Insider Intelligence. https://www.emarketer.com/content/global-ecommerce-2019

European Commission. (2020). *EURES - Labour market information - Stockholms län - European Commission*. https://ec.europa.eu/eures/main.jsp?countryId=SE&acro=lmi&showRegion=true&lang=en&mode=text®ionId=SE0&nuts2Code=SE01&nuts3Code=SE010&catId=2615

Eurostat. (2019). *Employed persons working from home as a percentage of the total employment, by sex, age and professional status*. https://appsso.eurostat.ec.europa.eu/nui/submitViewTableAction.do

Eurostat. (2020, July 20). *Impact of COVID-19 on main GDP aggregates including employment*
    [Eurostat]. Europa. https://ec.europa.eu/eurostat/statistics-
    explained/index.php?title=Impact_of_COVID-
    19_on_main_GDP_aggregates_including_employment

Gao, S., Rao, J., Kang, Y., Liang, Y., & Kruse, J. (2020). Mapping county-level mobility pattern
    changes in the United States in response to COVID-19. *ArXiv:2004.04544 [Physics, q-*
    *Bio]*. http://arxiv.org/abs/2004.04544

Gardshol, A. (2020). *E-barometern Q2 2020 - PostNord i samarbete med Svensk Digital Handel*
    *och HUI Research* (2020 Q2; e-barometer, p. 34). PostNord.

Google. (2020). *COVID-19 Community Mobility Report*. COVID-19 Community Mobility
    Report. https://www.google.com/covid19/mobility?hl=en

Hashem, T. N. (2020). Examining the Influence of COVID 19 Pandemic in Changing
    Customers' Orientation towards E-Shopping. *Modern Applied Science*, *14*(8), 59.
    https://doi.org/10.5539/mas.v14n8p59

Hoogendoorn-Lanser, S., Olde Kalter, M.-J., & Schaap, N. T. W. (2019). Impact of different
    shopping stages on shopping-related travel behaviour: analyses of the Netherlands
    Mobility Panel data. *Transportation*, *46*(2), 341–371. https://doi.org/10.1007/s11116-
    019-09993-7

IBM Corp. (2019). *IBM SPSS Statistics* (Version 26.0) [Windows]. IBM Corp.

Istat. (2020). *Statistiche Istat*. http://dati.istat.it/

Jay, J., Bor, J., Nsoesie, E., Lipson, S. K., Jones, D. K., Galea, S., & Raifman, J. (2020).
    *Neighborhood income and physical distancing during the COVID-19 pandemic in the*

*U.S.* [Preprint]. Infectious Diseases (except HIV/AIDS). https://doi.org/10.1101/2020.06.25.20139915

Kampf, G., Todt, D., Pfaender, S., & Steinmann, E. (2020). Persistence of coronaviruses on inanimate surfaces and their inactivation with biocidal agents. *Journal of Hospital Infection*, *104*(3), 246–251. https://doi.org/10.1016/j.jhin.2020.01.022

Kavanagh, N. M., Goel, R. R., & Venkataramani, A. S. (2020). *Association of County-Level Socioeconomic and Political Characteristics with Engagement in Social Distancing for COVID-19* [Preprint]. Health Policy. https://doi.org/10.1101/2020.04.06.20055632

Kraemer, M. U. G., Yang, C.-H., Gutierrez, B., Wu, C.-H., Klein, B., Pigott, D. M., Open COVID-19 Data Working Group†, du Plessis, L., Faria, N. R., Li, R., Hanage, W. P., Brownstein, J. S., Layan, M., Vespignani, A., Tian, H., Dye, C., Pybus, O. G., & Scarpino, S. V. (2020). The effect of human mobility and control measures on the COVID-19 epidemic in China. *Science*, *368*(6490), 493–497. https://doi.org/10.1126/science.abb4218

Kunzmann, K. R. (2020). Smart Cities After Covid-19: Ten Narratives. *DisP - The Planning Review*, *56*(2), 20–31. https://doi.org/10.1080/02513625.2020.1794120

Laurencin, C. T., & McClinton, A. (2020). The COVID-19 Pandemic: a Call to Action to Identify and Address Racial and Ethnic Disparities. *Journal of Racial and Ethnic Health Disparities*, *7*(3), 398–402. https://doi.org/10.1007/s40615-020-00756-0

Layer, R. M., Fosdick, B., Larremore, D. B., Bradshaw, M., & Doherty, P. (2020). *Case Study: Using Facebook Data to Monitor Adherence to Stay-at-home Orders in Colorado and Utah* [Preprint]. Public and Global Health. https://doi.org/10.1101/2020.06.04.20122093

Lee, R. J., Sener, I. N., Mokhtarian, P. L., & Handy, S. L. (2017). Relationships between the online and in-store shopping frequency of Davis, California residents. *Transportation Research Part A: Policy and Practice*, *100*, 40–52. https://doi.org/10.1016/j.tra.2017.03.001

Lisa Lockerd Maragakis. (2020, November 17). *Coronavirus Second Wave? Why Cases Increase*. Hopkins Medicine. https://www.hopkinsmedicine.org/health/conditions-and-diseases/coronavirus/first-and-second-waves-of-coronavirus

Lozzi, Rodrigues, Marcucci, Teoh, Gatta, & Pacelli. (2020). *Research for TRAN Committee – COVID-19 and urban mobility: impacts and perspectives*. European Parliament, Policy Department for Structural and Cohesion Policies, Brussels. https://www.unwto.org/impact-assessment-of-the-covid-19-outbreak-on-international-tourism

Malik, A. A., Couzens, C., & Omer, S. B. (2020). *COVID-19 related social distancing measures and reduction in city mobility* [Preprint]. Epidemiology. https://doi.org/10.1101/2020.03.30.20048090

Ministero del Lavoro e delle Politiche Sociali. (2020, maggio). *Smart working, le novità del Decreto Rilancio*. https://www.lavoro.gov.it/notizie/Pagine/Smart-working-le-novita-del-Decreto-Rilancio.aspx

Ministero della Salute. (2020). *Notizie*. http://www.salute.gov.it/portale/nuovocoronavirus/archivioNotizieNuovoCoronavirus.jsp

Mokhtarian, P. L. (1990). A typology of relationships between telecommunications and transportation. *Transportation Research Part A: General*, *24*(3), 231–242.

https://doi.org/10.1016/0191-2607(90)90060-J

Mokhtarian, P. L. (2002). Telecommunications and Travel: The Case for Complementarity. *Journal of Industrial Ecology*, *6*(2), 43–57. https://doi.org/10.1162/108819802763471771

Molloy, J., Tchervenkov, C., Hintermann, B., & Axhausen, K. W. (2020). Tracing the Sars-CoV-2 Impact: The First Month in Switzerland. *Findings*. https://doi.org/10.32866/001c.12903

Musselwhite, C., Avineri, E., & Susilo, Y. (2020). Editorial JTH 16 –The Coronavirus Disease COVID-19 and implications for transport and health. *Journal of Transport & Health*, *16*, 100853. https://doi.org/10.1016/j.jth.2020.100853

Nielsen, B. F., Sneppen, K., Simonsen, L., & Mathiesen, J. (2020). *Heterogeneity is essential for contact tracing* [Preprint]. Epidemiology. https://doi.org/10.1101/2020.06.05.20123141

NRF. (2019). *Fall 2019 Consumer View*. https://cdn.nrf.com/sites/default/files/2019-10/NRF%20Consumer%20View%20Fall%202019.pdf

Pani, A., Mishra, S., Golias, M., & Figliozzi, M. (2020). Evaluating public acceptance of autonomous delivery robots during COVID-19 pandemic. *Transportation Research Part D: Transport and Environment*, *89*, 102600. https://doi.org/10.1016/j.trd.2020.102600

Parady, G., Taniguchi, A., & Takami, K. (2020). Travel behavior changes during the COVID-19 pandemic in Japan: Analyzing the effects of risk perception and social influence on going-out self-restriction. *Transportation Research Interdisciplinary Perspectives*, *7*, 100181. https://doi.org/10.1016/j.trip.2020.100181

Pepe, E., Bajardi, P., Gauvin, L., Privitera, F., Lake, B., Cattuto, C., & Tizzoni, M. (2020). *COVID-19 outbreak response: a first assessment of mobility changes in Italy following national lockdown* [Preprint]. Infectious Diseases (except HIV/AIDS).

https://doi.org/10.1101/2020.03.22.20039933

Quilty, B. J., Diamond, C., Liu, Y., Gibbs, H., Russell, T. W., Jarvis, C. I., Prem, K., Pearson, C.
A. B., Clifford, S. J., Flasche, S., CMMID COVID-19 working group, Klepac, P., Eggo,
R. M., & Jit, M. (2020). *The effect of inter-city travel restrictions on geographical spread
of COVID-19: Evidence from Wuhan, China* [Preprint]. Epidemiology.
https://doi.org/10.1101/2020.04.16.20067504

Rahman, M. M., Thill, J.-C., & Paul, K. C. (2020). COVID-19 Pandemic Severity, Lockdown
Regimes, and People's Mobility: Early Evidence from 88 Countries. *Sustainability*,
*12*(21), 9101. https://doi.org/10.3390/su12219101

Regeringskansliet. (2020, April 9). *Decisions and guidelines in the Ministry of Health and Social
Affairs' policy areas to limit the spread of the COVID-19 virus* [Text]. Regeringskansliet;
Regeringen och Regeringskansliet. https://www.government.se/articles/2020/04/s-
decisions-and-guidelines-in-the-ministry-of-health-and-social-affairs-policy-areas-to-
limit-the-spread-of-the-covid-19-virusny-sida/

Rotem-Mindali, O., & Weltevreden, J. W. J. (2013). Transport effects of e-commerce: what can
be learned after years of research? *Transportation*, *40*(5), 867–885.
https://doi.org/10.1007/s11116-013-9457-6

Sabat, I., Neuman-Böhme, S., Varghese, N. E., Barros, P. P., Brouwer, W., van Exel, J.,
Schreyögg, J., & Stargardt, T. (2020). United but divided: policy responses and people's
perceptions in the EU during the COVID-19 outbreak. *Health Policy*,
S0168851020301639. https://doi.org/10.1016/j.healthpol.2020.06.009

Saphores, J.-D., & Xu, L. (2020). E-shopping changes and the state of E-grocery shopping in the US - Evidence from national travel and time use surveys. *Research in Transportation Economics*, 100864. https://doi.org/10.1016/j.retrec.2020.100864

SCB. (2019). *Arbetsmarknadssituationen för hela befolkningen 15-74 år, AKU 2019*. https://www.scb.se/contentassets/9d3fad266baf4bef96321252f80c7710/am0401_2019a01_sm_am12sm2001.pdf

SCB. (2020). *Gender statistics*. Statistiska Centralbyrån. http://www.scb.se/en/finding-statistics/statistics-by-subject-area/living-conditions/gender-statistics/gender-statistics/

Schmid, B. (2019). *Connecting Time-Use, Travel and Shopping Behavior: Results of a Multi-Stage Household Survey* [Doctoral Thesis, ETH Zurich]. https://doi.org/10.3929/ethz-b-000370588

Shi, K., De Vos, J., Yang, Y., & Witlox, F. (2019). Does e-shopping replace shopping trips? Empirical evidence from Chengdu, China. *Transportation Research Part A: Policy and Practice*, *122*, 21–33. https://doi.org/10.1016/j.tra.2019.01.027

Sunio, V., Schmöcker, J.-D., & Kim, J. (2018). Understanding the stages and pathways of travel behavior change induced by technology-based intervention among university students. *Transportation Research Part F: Traffic Psychology and Behaviour*, *59*, 98–114. https://doi.org/10.1016/j.trf.2018.08.017

Warren, M. S., & Skillman, S. W. (2020). Mobility Changes in Response to COVID-19. *ArXiv:2003.14228 [Cs]*. http://arxiv.org/abs/2003.14228

WHO. (2020, December 8). *Advice for the public on COVID-19 – World Health Organization*. https://www.who.int/emergencies/diseases/novel-coronavirus-2019/advice-for-public

World Economic Forum. (2019). *The global gender gap report 2020*.

https://www.weforum.org/reports/global-gender-gap-report-2020

Worldometer. (2020a). *Italy Population (2020) - Worldometer*.

https://www.worldometers.info/world-population/italy-population/

Worldometer. (2020b). *Sweden Population (2020) - Worldometer*.

https://www.worldometers.info/world-population/sweden-population/

Yechezkel, M., Weiss, A., Rejwan, I., Shahmoon, E., Ben Gal, S., & Yamin, D. (2020). *Human mobility and poverty as key factors in strategies against COVID-19* [Preprint].

Epidemiology. https://doi.org/10.1101/2020.06.04.20112417

# Appendix I - Survey structure

*Before COVID-19*

| Question | Activities | Options |
|---|---|---|
| **Q1: How often did you use to do these activities before the coronavirus outbreak?** | Commute to work or school | Less than once a month |
| | Travel to grocery stores | Once a month |
| | Purchase groceries online | Once every second week |
| | Travel to non-grocery shopping | Once a week |
| | Purchase non-groceries online | 2-4 times a week |
| | Order take away food | 5-7 times a week |
| | Eat out in restaurants, bars, cafes | More than daily |
| | Travel to visit friends and family | |
| | Go out for entertainment/hobbies | |
| | Perform physical activities | |
| **Q2: How did you use to travel to perform these activities before the coronavirus outbreak? If you use multimodal travelling, select the mode which covers the most distance and most commonly used.** | Commute to work or school | Car (driver) |
| | Travel to grocery stores | Car (passenger) |
| | Travel to non-grocery shopping | Motorcycle |
| | Eat out in restaurants, bars, cafes | Public transport or train |
| | Travel to visit friends and family | Bicycle or by foot |
| | Go out for entertainment/hobbies | Other |
| | Perform physical activities | N/A |
| **Q3: How long did it take to travel to these activities before the coronavirus outbreak? According to the mode selected above, estimate the average of one-way trip** | Commute to work or school | < 10min |
| | Travel to grocery stores | 10-30min |
| | Travel to non-grocery shopping | 30-60min |
| | Eat out in restaurants, bars, cafes | 1-2 hours |
| | Travel to visit friends and family | 3-5 hours |
| | Go out for entertainment/hobbies | > 5 hours |
| | Perform physical activities | N/A |
| **Q4: For how long did you use to use the internet connection for the following activities before the coronavirus outbreak? Answer considering an average in hours per week.** | For entertainment | < 1 h |
| | For personal videocall/call/chat | 1-2 h |
| | For work or study | 3-5 h |
| | For work or study meetings and calls | 6-10 h |
| | | 11-15 h |
| | | 15-20 h |
| | | 20-25 h |
| | | >25 h |
| **Q5: To what extent does the coronavirus outbreak influence your daily life?** | | 1 - not at all |
| | | 2 |
| | | 3 |
| | | 4 - a lot |

| Question | Activities | Options |
|---|---|---|
| Q6: How often do you do these activities now? | Commute to work or school | Less than once a month |
| | Travel to grocery stores | Once a month |
| | Purchase groceries online | Once every second week |
| | Travel to non-grocery shopping | Once a week |
| | Purchase non-groceries online | 2-4 times a week |
| | Order take away food | 5-7 times a week |
| | Eat out in restaurants, bars, cafes | More than daily |
| | Travel to visit friends and family | |
| | Go out for entertainment/hobbies | |
| | Perform physical activities | |
| Q7: How do you travel to perform these activities now? If you use multimodal travelling, select the mode which covers the most distance and most commonly used. | Commute to work or school | Car (driver) |
| | Travel to grocery stores | Car (passenger) |
| | Travel to non-grocery shopping | Motorcycle |
| | Eat out in restaurants, bars, cafes | Public transport or train |
| | Travel to visit friends and family | Bicycle or by foot |
| | Go out for entertainment/hobbies | Other |
| | Perform physical activities | N/A |
| Q8: How long does it take to travel to these activities now? According to the mode selected above, estimate the average of one-way trip | Commute to work or school | < 10min |
| | Travel to grocery stores | 10-30min |
| | Travel to non-grocery shopping | 30-60min |
| | Eat out in restaurants, bars, cafes | 1-2 hours |
| | Travel to visit friends and family | 3-5 hours |
| | Go out for entertainment/hobbies | > 5 hours |
| | Perform physical activities | N/A |
| Q9: For how long do you use the internet connection for the following activities now? Answer considering an average in hours per week. | For entertainment | < 1 h |
| | For personal videocall/call/chat | 1-2 h |
| | For work or study | 3-5 h |
| | For work or study meetings and calls | 6-10 h |
| | | 11-15 h |
| | | 15-20 h |
| | | 20-25 h |
| | | >25 h |

*Online shopping behaviours*

| Question | Activities | Options |
|---|---|---|
| Q10: Before the coronavirus outbreak, how much of your shopping was online? | Grocery shopping | 0% [online] |
| | Other shopping | 10%-30% |
| | | 40%-60% |
| | | 70%-90% |
| | | 100% [online] |
| Q11: How much of your shopping during the coronavirus outbreak is online? | Grocery shopping | 0% [online] |
| | Other shopping | 10%-30% |
| | | 40%-60% |
| | | 70%-90% |
| | | 100% [online] |
| Q12: Which kind of items do you usually shop during the coronavirus outbreak (not considering groceries)? [Multiple choice] | | Clothes |
| | | Hobbies related items (sport, art, music equipment, technology, stationery, books..) |
| | | Items for the house, garden |
| | | Work related items |
| | | Others [Specify] |

*Perceived safety and likelihood of keeping the new habits*

| Question | Activities | Options |
|---|---|---|
| Q13: During the coronavirus outbreak, how safe do you feel while engaging in the following activities, considering possible precautions that you take? | Travelling by public transport or train | 1 - not safe at all |
| | Travelling by car | 2 |
| | Visiting stores | 3 |
| | Being at the workplace or school | 4 - very safe |
| | Going to restaurants, pubs and cafes | |
| | Going to the gym | |
| | Spending time outside | |
| | Receiving home deliveries | |
| Q14: If you have changed your behaviour since the coronavirus outbreak, how likely are you to keep your new habits when the threat from the virus is removed? | Travel and commuting | 1 - not likely |
| | Grocery shopping | 2 |
| | Shopping | 3 |
| | Work or study | 4 - very likely |
| | Handle meetings at work or school | no change |
| | Free time | |
| | Physical activities | |

*Demographics*

| Question | Options | Number |
|---|---|---|
| Q15: Where do you currently live? | List of UN recognized states | |
| Q16: What is your postal code? | Open question | |
| Q17: What is your country of origin? | List of UN recognized states | |
| Q18: What gender do you identify with? | Female | |
| | Male | |
| | Other [Specify] | |
| Q19: How many people in these age groups live in your household? | Children (<18 years old) | None |
| | Adults | 1-3 |
| | Elderly (>65 years old) | More than 3 |
| Q20: What is your highest level of education? | No schooling completed | |
| | High school graduate, diploma or equivalent | |
| | Trade/technical/vocal training | |
| | Bachelor's degree | |
| | Master's degree | |
| | Professional degree | |
| | Doctorate degree | |
| | Other [Specify] | |
| Q21: What is your main employment? | Employee | |
| | Self-employed | |
| | Housewife/houseman | |
| | Student | |
| | Part time worker | |
| | Volunteering | |
| | Military | |
| | Retired | |
| | Unemployed | |
| | Other [Specify] | |
| Q22: Do you have any comments to add? | Open question | |